\documentclass[aps,prd,notitlepage,showpacs,nofootinbib,superscriptaddress]{revtex4-1}

\usepackage{graphicx}
\usepackage[utf8]{inputenc} 

\usepackage{amsmath}
\usepackage{amssymb}
\usepackage{float}
\usepackage{comment}
\usepackage{slashed}
\usepackage[normalem]{ulem}

\usepackage{caption}
\usepackage{subcaption}

\usepackage{amsmath}
\usepackage{amssymb}

\usepackage[usenames,dvipsnames]{color}
\usepackage[colorinlistoftodos]{todonotes}
\usepackage[colorlinks=true,citecolor=darkred,urlcolor=darkred, pdfborder={0 0 0}]{hyperref}
\usepackage[normalem]{ulem}

\definecolor{darkred}{rgb}{0.6,0,0}
\usepackage[colorinlistoftodos]{todonotes}

\definecolor{linkcolor}{rgb}{0,0,0.5}

\newcommand {\ignore}[1]{}

\def\gsim{\raise0.3ex\hbox{$\;>$\kern-0.75em\raise-1.1ex\hbox{$\sim\;$}}}
\def\lsim{\raise0.3ex\hbox{$\;<$\kern-0.75em\raise-1.1ex\hbox{$\sim\;$}}}

\def\SM{$\mathrm{SU(3)_c \otimes SU(2)_L \otimes U(1)_Y}$ }

\usepackage{soul}

\definecolor{mightnightblue}{RGB}{25,25,112}

\definecolor{brown}{rgb}{0.59, 0.29, 0.0}

\newcommand {\black} {\color{black}}

\def\vev#1{\left\langle #1\right\rangle}
\def\SM{$\mathrm{SU(3)_c \otimes SU(2)_L \otimes U(1)_Y}$ }
\def\21{$\mathrm{SU(2)_L \otimes U(1)_Y}$}

\newcommand{\AddrAHEP}{%
  AHEP Group, Institut de F\'{i}sica Corpuscular --
  CSIC/Universitat de Val\`{e}ncia, Parc Cient\'ific de Paterna.\\
 C/ Catedr\'atico Jos\'e Beltr\'an, 2 E-46980 Paterna (Valencia) - SPAIN}
 
 \newcommand{\AddrUSM}{Universidad T\'{e}cnica Federico Santa Mar\'{\i}a, Casilla 110-V, Valpara\'{\i}so, Chile}
 
\newcommand{\CCTVAL}{Centro Cient\'{\i}fico-Tecnol\'ogico de Valpara\'{\i}so, Casilla 110-V, Valpara\'{\i}so, Chile}

\newcommand{\SAPHIR}{Millennium Institute for Subatomic Physics at High-Energy Frontier (SAPHIR), Fern\'andez Concha 700, Santiago, Chile}

\usepackage{xcolor,cancel}

\begin{document}


\title{\color{BrickRed} Linear seesaw mechanism from dark sector}

\author{A. E. C\'{a}rcamo Hern\'{a}ndez}\email{antonio.carcamo@usm.cl}
\affiliation{\AddrUSM} \affiliation{\CCTVAL} \affiliation{\SAPHIR}

\author{Vishnudath K. N.}
\email{vishnudath.neelakand@usm.cl}
\affiliation{\AddrUSM}

\author{Jos\'{e} W. F. Valle}\email{valle@ific.uv.es}
\affiliation{\AddrAHEP}

\begin{abstract} 
\vspace{0.5cm} 
We propose a minimal model where a dark sector seeds neutrino mass generation radiatively within the linear seesaw mechanism.
Neutrino masses are calculable, since tree-level contributions are forbidden by symmetry.
{They arise from spontaneous lepton number violation by a small Higgs triplet vacuum expectation value.}
Lepton flavour violating processes e.g. $\mu \to e\gamma$ can be sizeable, despite the tiny neutrino masses.
We comment also on dark-matter and collider implications.

\end{abstract}  

\maketitle
\noindent

\section{Introduction} 
\label{Sect:intro}
Two solid indications for new physics beyond the Standard Model (SM) are the existence of neutrino masses~\cite{McDonald:2016ixn,Kajita:2016cak} and dark matter~\cite{Bertone:2004pz}.
There are many ways to induce neutrino masses, and at the moment we do not know which one is nature's choice.
There are also many options to add new electrically neutral fermions and/or scalars to the SM so as to provide a viable dark matter (DM) candidate.
Typically the latter is made stable through the imposition of an adequate ``dark parity'' symmetry.

In most of these SM extensions there is no relation between dark matter and neutrino mass generation.
There has been a recent suggestion that neutrino mass generation proceeds \textit{a la seesaw} within the \SM framework~\cite{Schechter:1980gr,Schechter:1981cv},
seeded by a dark sector~\cite{Ma:2009gu,Law:2012mj,Ahriche:2016acx,Rojas:2018wym,Mandal:2019oth,Abada:2021yot}~\footnote{
    Low-scale seesaw schemes have been investigated extensively in recent years~\cite{ CarcamoHernandez:2019eme,CarcamoHernandez:2020pnh,Hernandez:2021mxo}.
    In particular, radiative constructions implementing, e.g., supersymmetry, extended gauge and/or family symmetries have been proposed,
    see~\cite{Bazzocchi:2009kc,CarcamoHernandez:2017owh,CarcamoHernandez:2017kra,CarcamoHernandez:2018hst,CarcamoHernandez:2019lhv,CarcamoHernandez:2019vih,
      CarcamoHernandez:2019pmy,Hernandez:2021uxx,Hernandez:2021xet}. }.
This way neutrino mass generation becomes intimately connected with dark matter physics.

In this paper we suggest that neutrino masses are seeded by a dark sector within the context of the linear seesaw
mechanism~\cite{Akhmedov:1995ip,Akhmedov:1995vm,Malinsky:2005bi}.
Several SM extensions implementing the linear seesaw mechanism have been used recently to approach the flavor problem~\cite{CarcamoHernandez:2017cwi,CarcamoHernandez:2019iwh,CarcamoHernandez:2021tlv}.
  Here we focus on the dark matter issue,
  proposing a dark-seeded linear seesaw mechanism, as an alternative to the dark-seeded extension of the inverse seesaw~\cite{Mohapatra:1986bd,Gonzalez-Garcia:1988okv}.
This new realization makes use of the simplest template structure shared by all low-scale seesaw schemes, and employs a very simple dark sector.
The latter consists of one SM doublet and a singlet dark scalar, while for dark fermions we employ three SM singlet two-component Majorana fermions.
The Higgs sector contains, besides the SM doublet, a {complex} isotriplet involved in seeding neutrino mass generation.

Many phenomenological implications of our proposal are also expected in generic linear seesaw setups~\cite{Batra:2023ssq,Batra:2023mds}.
Besides these, we have a Weakly Interacting Massive Particle (WIMP) dark-matter candidate, that can be identified with the lightest electrically neutral dark particle (LDP).
Rather than being associated to supersymmetry~\cite{Jungman:1995df}, WIMP dark-matter emerges here as a neutrino mass mediator,
in a manner distinct from scotogenic approaches~\cite{Ma:2006km,Hirsch:2013ola},
and also inequivalent to the dark inverse-seesaw realization~\cite{Mandal:2019oth}.
We examine the most salient phenomenological implications of the dark linear seesaw mechanism,
concerning charged lepton flavor violation, and comment also on dark matter and collider physics implications.

\section{The model}
\label{sec:model}

\begin{figure}[tbh]
\centering
\includegraphics[width=.5\textwidth,height=4cm]{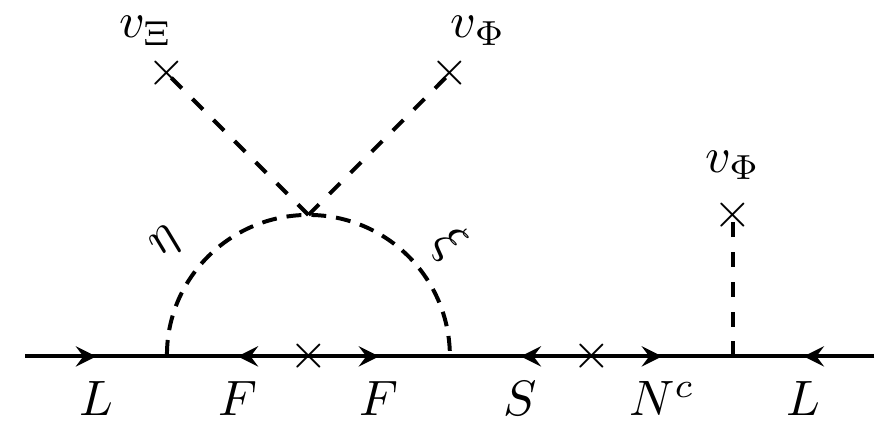}
\caption{Feynman diagram for neutrino mass generation in dark linear seesaw mechanism (plus symmetrization).}
\label{fig:Neutrinoloopdiagram}
\end{figure}

Our proposed model can be seen as a minimal extension of the inert doublet model where the linear seesaw mechanism producing the tiny neutrino masses
is implemented at the one-loop level, seeded by the dark sector.
The SM lepton sector is enlarged by the inclusion of the neutral leptons $N _{i}^{c}$ and $S_{i}$ ($i=1,2,3$), characteristic of low-scale seesaw schemes.
The dark sector contains three copies of SM singlet two-component Majorana fermions $F_{i}$, plus a SM doublet dark scalar $\eta $, and a dark gauge singlet $\xi$.
These dark scalars $\eta$ and $\xi$ and the dark fermions $F_{i}$ seed linear-seesaw neutrino mass generation as seen in Fig~\ref{fig:Neutrinoloopdiagram}.
The $\mathrm{SU(3)_c \otimes SU(2)_L \otimes U(1)_Y}$ gauge symmetry is supplemented by the inclusion of the global 
$U\left( 1\right)_{\cal L} $ lepton number symmetry, which spontaneously breaks to a preserved $\mathcal{Z}_{2}$ symmetry.
 This remnant symmetry, which can be identified as the matter parity $(-1)^{3 \mathcal{B} + \mathcal{L} + 2s}$ with $\mathcal B$, $\mathcal L$ and $s$ being the baryon,
  lepton and spin quantum numbers respectively, ensures the stability of the dark matter candidate, as well as the radiative nature of neutrino mass generation through the linear seesaw mechanism.

The scalar sector of our model also requires Higgs bosons to drive spontaneous breaking of the gauge and global symmetries.
Besides the SM doublet $\Phi$, we include a {complex scalar isotriplet $\Xi$} whose vacuum expectation value (VEV)
is restricted by precision electroweak measurements, i.e. the $\rho$ parameter~\cite{ParticleDataGroup:2022pth}.
 The leptons and scalars of the model and their transformation properties under the $\mathrm{SU(3)_c \otimes SU(2)_L \otimes U(1)_Y}$
gauge symmetry, the global lepton number symmetry, and the remnant $Z_2$ symmetry are given in Table.~\ref{tab:MatterModel}.

\begin{table}[h] 
  \setlength\tabcolsep{0.25cm}
  \centering%
\begin{tabular}{|c||c|c|c|c||c|c||c||c|c|}
\hline
& $L_{i}$ & $l_{i}^{c}$ & $N_{i}^{c}$ & $S_{i}$ & $\Phi $ & $\Xi$ &  $F_{i}$   & $\eta $ & $\xi $ \\ \hline\hline
$SU(2)_{L}\times U(1)_{Y}$ & $(2,-\frac{1}{2})$ & $(1,1)$ & $(1,0)$ & $(1,0)$ & $(2,\frac{1}{2})$ & $(3,0)$ & $(1,0)$
 &  $(2,\frac{1}{2})$ & $(1,0)$ \\ 
$U\left( 1\right) _{\cal L }$ & $1$ & $-1$ & $-1$ & $1$ & $0$ & $2$ & $0$ & $-1$ & {$-1$} \\ 
$(-1)^{3 \mathcal{B} + \mathcal{L} + 2s}$ & $1$ & $1$ & $1$ & $1$ & $1$ & $1$ & $-1$ & $-1$ & $-1$\\ \hline
\end{tabular}%
 \caption{Fields and their quantum numbers. All fermions come in three copies, $i=1,2,3$.}
\label{tab:MatterModel}
\end{table}

Notice that the leptons have the conventional lepton number assignment characteristic of low-scale seesaw schemes.
Together with the dark scalars, the new Majorana neutral fermions $F_{i}$ play a key role in seeding non-zero neutrino masses. 
Except for the SM scalar doublet $\Phi$, all scalars carry non-zero lepton number.
The relevant neutrino Yukawa couplings and mass terms invariant under these symmetries are,
\begin{align}
  \label{eq:Yuk}
-\mathcal{L}_{Y}^{\left( \nu \right)
}=& \sum_{i,j=1}^{3}Y_{ij}^{\left( \Phi \right)
}L_{i}^{T}CN_{j}^{c}\Phi +\sum_{i,j=1}^{3}Y_{ij}^{\left( \eta
\right) }L_{i}^{T}CF_{j}\eta +\sum_{i,j=1}^{3}Y_{ij}^{\left( \xi
\right) }S_{i}^T C F_{j}\xi \notag \\& +\sum_{i=1}^{3}\left( m_{F}\right)
_{i}F_{i}^T C F_{i}+\sum_{i,j=1}^{3}M_{ij}{N_{i}^{c}}^TCS_{j}\,{ + \sum_{i,j=1}^{3} {Y^\prime}^{(\xi)}_{ij}F_i^T C {N^c}_j \xi^*} +  H.c.
\end{align}

The scalar potential contains,  
\begin{align}
  \label{eq:V1}
\mathcal{V}_{\left( s\right) }=& -\mu _{\Phi }^{2}(\Phi ^{\dagger }\Phi
\,)-\mu _{\Xi }^{2}Tr(\Xi ^{\dagger }\Xi )  + \mu _{\eta }^{2}(\eta ^{\dagger }\eta )+\mu _{\xi }^{2}(\xi ^{\ast }\xi
) ~{~+A_{\Phi }(\Phi ^{\dagger }\Xi \Phi + \Phi ^{\dagger }\Xi^\dagger \Phi)} \notag  \\
& +\lambda _{1}(\Phi ^{\dagger }\Phi \,)^{2}+\lambda _{2}(\eta ^{\dagger
}\eta )^{2}+\lambda _{3}(\xi ^{\ast }\xi )^{2}+\lambda _{4}\left[ Tr(\Xi
^{\dagger }\Xi )\right] ^{2}+\lambda _{5}Tr\left[ (\Xi ^{\dagger }\Xi )^{2}%
\right]  \notag \\
& +\lambda _{6}(\Phi ^{\dagger }\Phi \,)(\eta ^{\dagger }\eta )+\lambda
_{7}(\Phi ^{\dagger }\eta )(\eta ^{\dagger }\Phi \,)+\lambda _{8}(\Phi
^{\dagger }\Phi \,)Tr(\Xi ^{\dagger }\Xi )+\lambda _{9}\Phi ^{\dagger }\Xi
\Xi ^{\dagger }\Phi  \notag \\
& +\lambda _{10}(\Phi ^{\dagger }\Phi \,)(\xi ^{\ast }\xi )+\lambda
_{11}(\eta ^{\dagger }\eta )Tr(\Xi ^{\dagger }\Xi )+\lambda _{12}\eta
^{\dagger }\Xi \Xi ^{\dagger }\eta +\lambda _{13}(\eta ^{\dagger }\eta )(\xi
^{\ast }\xi )  \notag \\
& +\lambda _{14}(\xi ^{\ast }\xi )Tr(\Xi ^{\dagger }\Xi )+ {\lambda
_{15}\left(\eta^{\dag }\Xi^{\ast }\Phi\xi ^{\ast }+h.c\right)} .
\end{align}

The U(1)$_{\cal L }$ symmetry is broken by the VEV of the neutral part of $\Xi$.
{The presence of the trilinear term $A_\Phi$ in Eq.~(\ref{eq:V1}) also breaks the global U(1)$_{\cal L }$ symmetry of Eqn.~(\ref{eq:Yuk}), explicitly but softly.}
Dark matter stability is ensured by the remnant unbroken $\mathcal{Z}_{2}$ symmetry preserved after the breaking of the $U\left( 1\right) _{\cal L }$ symmetry.
To ensure this we require that the $\mathcal{Z}_{2}$-odd scalars $\eta $ and $\xi$ do not acquire vacuum expectation values. 

The scalar fields $\Phi $, $\Xi $, $\eta $ and $\xi $ can be written as follows, 
\begin{eqnarray*}
\Phi &=&%
\begin{pmatrix}
\phi ^{+} \\ 
\frac{v_{\Phi }+\phi _{R}^{0}+i\phi _{I}^{0}}{\sqrt{2}}%
\end{pmatrix}%
,\hspace{3.3cm}\eta =%
\begin{pmatrix}
\eta ^{+} \\ 
\frac{\eta _{R}^{0}+i\eta _{I}^{0}}{\sqrt{2}}%
\end{pmatrix}%
, \\
  \Xi &=&  {\left( 
\begin{array}{cc}
\frac{v_{\Xi } + \Xi ^{0}_R +  i\Xi ^{0}_I}{\sqrt{2}} & \Xi_1^{+} \\ 
\Xi_2^{-} & -\frac{v_{\Xi } + \Xi ^{0}_R +  i\Xi ^{0}_I}{\sqrt{2}}
\end{array}%
\right) } ,\hspace{1cm}\xi =\frac{\xi _{R}+i\xi _{I}}{\sqrt{2}}.
\end{eqnarray*}

{We have two charged Higgs scalars $\Xi_1^{\pm}$ and $\Xi_2^{\pm}$ with mass-squared given as, 
\begin{equation}
m_{\Xi_{1,2}^{\pm       }}^2 = \frac{\sqrt{2} A_\Phi (v_\Phi^2 + 4 v_\Xi^2) \mp \sqrt{32 A_\Phi^2 v_\Xi^4 + v_\Phi^2 v_\Xi^2 (v_\Phi^2 + 8 v_\Xi^2)\lambda_9^2}}{4 v_\Xi}, 
 \end{equation} }
and a charged dark scalar $\eta^{\pm}$ with mass-squared given as,
\begin{equation}
m_{\eta^\pm}^2 =  \frac{1}{2} v_\Xi^2 (2 \lambda_{11} +\lambda_{12}) + 
\frac{1}{2} v_\Phi^2 \lambda_6  +
\mu_\eta^2.
\end{equation}
{We also note that, in the presence of the cubic term $A_\Phi$ the two charged components of the triplet scalar will have an adequate mass-squared term.}

Electroweak symmetry breaking is driven mainly by the VEV of $\Phi $.
The resulting mass squared matrices for the CP-even neutral Higgs scalars are given as,
\begin{equation} 
M^2_{\phi_{R}^{0}~\Xi_R^{0}} = \left( 
\begin{array}{cc}
2 \lambda_1 v_\Phi^2 &  v_\Phi(-\sqrt{2}A_\Phi +  v_\Xi (2
\lambda_8 + \lambda_9)) \\ 
 v_\Phi(-\sqrt{2}A_\Phi +  v_\Xi (2
\lambda_8 + \lambda_9)) & 
\frac{A_\Phi v_\Phi^2}{\sqrt{2}v_\Xi} + 4 v_\Xi^2 (2 \lambda_4 + \lambda_5)%
\end{array}%
\right) ,
\end{equation}
while the corresponding neutral dark scalar mass squared matrices are given as,
\begin{equation}
M^2_{\eta_{R}^{0}~\xi_{R}} = \left( 
\begin{array}{cc} \frac{v_\Xi^2}{2} ( 2 \lambda_{11} +
\lambda_{12}) + \frac{v_\Phi^2}{2}(\lambda_6 + \lambda_7) + \mu_\eta^2 & -%
\frac{1}{2}\lambda_{15} v_\Phi v_\Xi \\ 
-\frac{1}{2}\lambda_{15} v_\Phi v_\Xi & \frac{1}{2}\lambda_{10} v_\Phi^2  +
\lambda_{14} v_\Xi^2  + \mu_\xi^2 
\end{array}
\right) ,  \label{thetas}
\end{equation}
\begin{equation}
M^2_{\eta_{I}^{0}~\xi_{I}} = \left( 
\begin{array}{cc} \frac{v_\Xi^2}{2} ( 2 \lambda_{11} +
\lambda_{12}) + \frac{v_\Phi^2}{2}(\lambda_6 + \lambda_7) + \mu_\eta^2 & 
 \frac{1}{2}\lambda_{15} v_\Phi v_\Xi \\ 
 \frac{1}{2}\lambda_{15} v_\Phi v_\Xi & \frac{1}{2}\lambda_{10} v_\Phi^2  +
\lambda_{14} v_\Xi^2  + \mu_\xi^2 
\end{array}
\right) .  \label{thetaa}
\end{equation}
{There is also a CP-odd scalar coming from the imaginary part of the neutral component of $\Xi$, whose mass is given as,
\begin{equation} 
m_{{\Xi^0}_I}^2 = \frac{A_\Phi v_\Phi^2}{\sqrt{2}v_\Xi}.
\end{equation}}
Thus, one sees that the physical scalar spectrum includes four CP-even scalars:
two neutral Higgs $H_1$ and $H_2$ arising from the mixing of $\Xi^{0}_R$ and $\phi_{R}^{0}$, and containing the 125~GeV SM Higgs boson~\cite{ATLAS:2012yve,CMS:2012qbp},
plus two dark neutral scalars $D_1$ and $D_2$ arising from the mixing of $\eta_{R}^{0}$ and $\xi_{R}$.
Moreover, we have two dark neutral CP-odd scalars, $D_{A_1}$ and $D_{A_2}$ arising from the mixing of $\eta_{I}^{0}$ and $\xi_{I}$ {and another CP-odd scalar corresponding to ${\Xi^0}_I$.}
The doublet-singlet mixing angles in these matrices are expected to be naturally small, thanks to the limit $v_\Xi \lsim 3$GeV from the $\rho$ parameter.\\

{Note also that, thanks to the cubic lepton number soft-breaking term $A_\Phi$ present in Eq.~(\ref{eq:V1}), all physical scalars are massive.
This avoids the existence of a Majoron~\cite{Schechter:1981cv,Chikashige:1980ui}, a physical Nambu-Goldstone boson associated to spontaneous lepton number
violation. This gets an adequately large mass from the explicit $A_\Phi$-induced breaking of lepton number.
An alternative full-fledged Majoron scheme can also be implemented, along the lines of Ref.~\cite{Fontes:2019uld}.
However we do not pursue such extension here, as it is not essential for the neutrino mass generation.
}

\vskip .5cm
We now turn to the radiative neutrino mass generation via linear seesaw mechanism. 
The lepton Yukawa interactions yield the following neutrino mass terms, 
\begin{equation}
-\mathcal{L}_{mass}^{\left( \nu \right) }=\frac{1}{2}\left( 
\begin{array}{ccc}
\nu^T & {N^{c}}^T & S^T%
\end{array}%
\right) M_{\nu }C\left( 
\begin{array}{c}
\nu \\ 
N^{c} \\ 
S%
\end{array}%
\right) +H.c.,  \label{Lnu}
\end{equation}%
where the neutrino mass matrix $M_{\nu }$ is given as, 
\begin{equation}
M_{\nu }=\left( 
\begin{array}{ccc}
0_{3\times 3} & m_{D} & \varepsilon \\ 
m_{D}^{T} & 0_{3\times 3} & M \\ 
\varepsilon ^{T} & M & 0_{3\times 3}%
\end{array}%
\right) .  \label{M3}
\end{equation}%
Here the submatrix $M$ is a bare mass, and $m_{D}$ is generated at tree-level after electroweak symmetry breaking,
\begin{equation}
  \label{eq:MD}
  m_{D} = Y^{\left( \Phi \right) }\frac{v_{\Phi }}{\sqrt{2}}.
\end{equation}
In contrast, the small entry $\varepsilon$ arises radiatively, mediated by the one-loop level exchange of the dark fermions and scalars. 
The one-loop level Feynman diagram is given in Fig.~\ref{fig:Neutrinoloopdiagram} and the resulting submatrix $\varepsilon $ is given as,
\begin{equation}
\varepsilon _{ij} = \sum_{k=1}^{3}\frac{Y_{ik}^{\left( \eta \right)
}Y_{jk}^{\left( \xi \right) }M_{F_k}}{16\pi ^{2}}\left\{ \left[ f\left(
m_{D_{1}}^{2},m_{F_{k}}^{2}\right) - f\left(
m_{D_{2}}^{2},m_{F_{k}}^{2}\right) \right] \sin 2\theta_D - \left[ f\left(
m_{D_{A_1}}^2,m_{F_{k}}^{2}\right) -f\left(
m_{D_{A_2}}^2,m_{F_{k}}^{2}\right) \right] \sin 2\theta_{D_A}\right\} ,
\end{equation}
where $f\left( m_{1},m_{2}\right) $ is the function defined as,
\begin{equation}
f\left( m_{1},m_{2}\right) =\frac{m_{1}^{2}}{m_{1}^{2}-m_{2}^{2}}\ln \left( 
\frac{m_{1}^{2}}{m_{2}^{2}}\right) .
\end{equation}%
Here $m_{D_1}$ and $m_{D_2}$ are the masses of the physical CP even dark scalars, whereas $m_{D_{A_1}}$ and $m_{D_{A_2}}$ are those that of the dark pseudoscalars.
Their mixing matrices are defined as, 
\begin{equation}
\left( 
\begin{array}{c}
D_{1} \\ 
D_{2}%
\end{array}%
\right) =\left( 
\begin{array}{cc}
\cos \theta _{D} & \sin \theta _{D} \\ 
-\sin \theta _{D} & \cos \theta _{D}%
\end{array}%
\right) \left( 
\begin{array}{c}
\eta _{R} \\ 
\xi _{R}%
\end{array}%
\right) ,\hspace{1cm}\left( 
\begin{array}{c}
D_{A_1} \\ 
D_{A_2}%
\end{array}%
\right) =\left( 
\begin{array}{cc}
\cos \theta _{D_A} & \sin \theta _{D_A} \\ 
-\sin \theta _{D_A} & \cos \theta _{D_A}%
\end{array}%
\right) \left( 
\begin{array}{c}
\eta _{I} \\ 
\xi _{I}%
\end{array}%
\right) .
\end{equation}
where the small doublet-singlet mixing angles $\theta _{D}$ and $\theta _{D_A}$ are obtained by diagonalizing Eqns. (\ref{thetas}) and (\ref{thetaa}), respectively,
{from which one sees $\theta_{D_A}= - \theta_A$.}

Given a non-zero submatrix $\varepsilon$, the light active neutrino masses arise from the linear seesaw
mechanism~\cite{Akhmedov:1995ip,Akhmedov:1995vm,Malinsky:2005bi}, so that the resulting active-neutrino mass matrix has the form,
\begin{equation}
m_{\rm light}=-\left[ m_{D}M^{-1}\varepsilon ^{T}+\varepsilon M^{-1}m_{D}^{T}%
\right] .
\end{equation}
One sees that {spontaneous lepton number violation through $v_\Xi$} provides a radiative seed for light neutrino mass generation that proceeds \textit{a la seesaw}.
The smallness of the light neutrino masses is ascribed to the smallness of loop-suppressed $\varepsilon$ as well as the small,
but not necessarily negligible, $\frac{m_{D}}{M}$ ratio.
It is worth mentioning that small neutrino masses are symmetry-protected, making the model natural in t'Hooft's sense.

On the other hand, as in all low-scale seesaw models, our heavy mediator neutrino sector consists of three pairs of
quasi-Dirac~\cite{Valle:1982yw,Anamiati:2016uxp,Arbelaez:2021chf} heavy neutrinos, whose mass matrices are given as,
\begin{eqnarray}
M_{N^{-}} &=&-M-m_{D}^{T}m_{D}M^{-1}+\frac{1}{2}\left[ m_{D}M^{-1}%
\varepsilon ^{T}+\varepsilon M^{-1}m_{D}^{T}\right] , \\
M_{N^{+}} &=&M+m_{D}^{T}m_{D}M^{-1}+\frac{1}{2}\left[ m_{D}M^{-1}\varepsilon
^{T}+\varepsilon M^{-1}m_{D}^{T}\right] .
\end{eqnarray}  

\section{Phenomenology}
\label{sec:phenomenology} 

\subsection{Charged lepton flavor violation}

In this section we discuss the implications of our model for charged lepton flavor violation (cLFV). In particular, we focus on the radiative decays
$\ell_i\to\ell_j\gamma$, the most sensitive of which is the decay $\mu \to e\gamma $.
A key conceptual feature of low-scale seesaw, such as our proposed dark linear seesaw-scheme, is that leptonic flavour and CP can be violated even in the limit of massless
neutrinos~\cite{Bernabeu:1987gr,Langacker:1988up,Branco:1989bn,Rius:1989gk}.
That means that the cLFV rates are not suppressed by the small neutrino masses, and can therefore be sizeable.
\begin{figure*}[tbh]
\begin{center}
  \includegraphics[scale=0.3]{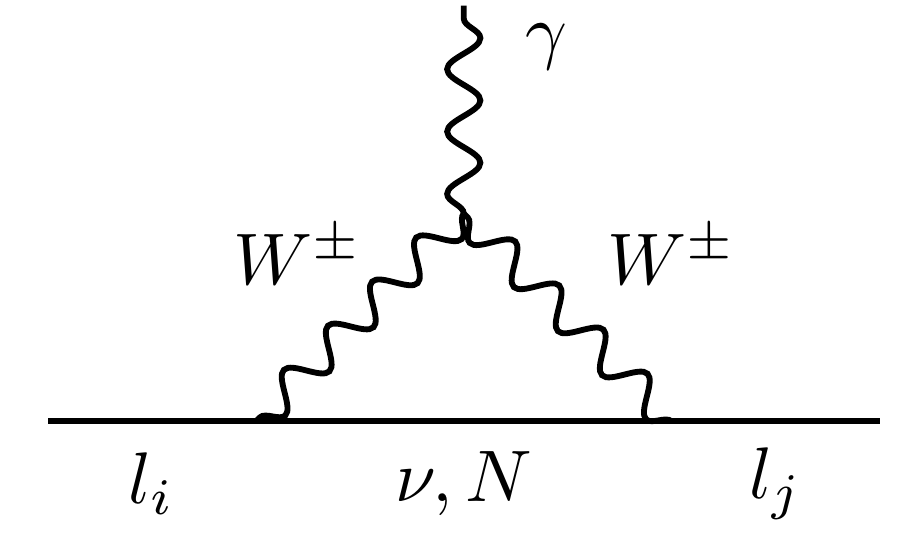}
  \includegraphics[scale=0.3]{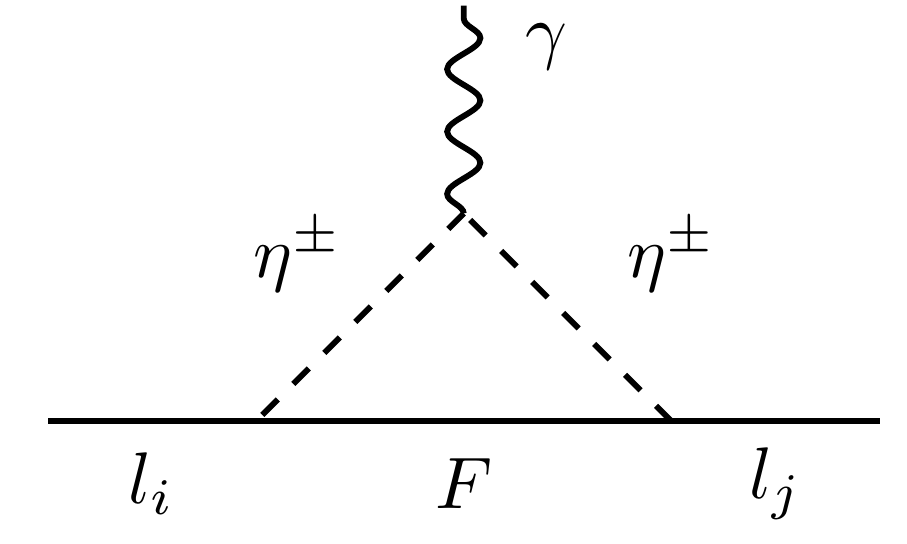}
\end{center}
\caption{
  Feynman diagrams that contribute to $l_{i}\to l_{j}\gamma$ processes.
  The left diagram shows the charged-current contribution, whereas the one in the right shows the dark-sector contribution.}
\label{fig:figmuegFeyn}
\end{figure*}

In the same spirit as~\cite{Casas:2001sr} one can give an approximate expression for extracting the ``Dirac'' submatrix in terms
of the measured oscillation parameters as follows~\cite{Forero:2011pc,Cordero-Carrion:2019qtu},
\begin{equation}
m_{D} = U_{lep} \sqrt{m_\nu} A^{T} \sqrt{m_\nu} U_{lep}^{T}\left(
\varepsilon ^{T}\right) ^{-1}M^{T} , ~~~~~~~\textrm{with} ~~~ A =\left( 
\begin{array}{ccc}
\frac{1}{2} & a & b \\ 
-a & \frac{1}{2} & c \\ 
-b & -c & \frac{1}{2}%
\end{array}
\right) , \label{mdeqn}
\end{equation}
where $a$, $b$ and $c$  are taken to be real numbers, $m_\nu = {\rm diag}\left( {m_{1}},{m_{2}},{m_{3}}\right)$ is given by the light neutrino masses, and $U_{lep}$ is approximately the lepton mixing matrix determined
in oscillation experiments~\cite{deSalas:2020pgw}. We assume the basis in which the charged lepton mass matrix is diagonal.
From this expression it is clear that, given $m_\nu$ from oscillation measurements, one can fix $m_{D}$ in accordance with
  the corresponding mass scale ratio $\varepsilon / M$ so as to fit the oscillations. Likewise, concerning mixing parameters,
  one can choose the off-diagonal entries of $m_{D}$ to match the measured solar, atmospheric and reactor mixing angles.
  In fact, taking as a simple ansatz $M$ and $\varepsilon$ diagonal and proportional to identity, one sees that $m_{Dij}$
  can be chosen so as to fit the observed mixing angles.

The $\mu \to e\gamma $ decay amplitude involves the Feynman diagrams in Fig.~\ref{fig:figmuegFeyn}.
There are two types of contributions, i.e. the charged-current (CC) contribution (left diagram) and a contribution arising from the dark sector (right diagram).
In order to determine the CC contribution the key ingredient is the full lepton mixing matrix, which has a rectangular form~\cite{Schechter:1980gr,Schechter:1981cv}.
Such rectangular mixing matrix describes not only the CC couplings of the light neutrinos, which gives a sizeable contribution due to
the effective unitarity violation, of order $\left(\frac{m_{D}}{M}\right)^2$,
but also the heavy mediator neutrino admixture in the left-handed CC weak interaction, of order $\left(\frac{m_{D}}{M}\right)$. 
We find that the CC light-neutrino contribution to the $\mu \to e\gamma $ decay can be sizeable, thanks to effective unitarity violation of the relevant coupling sub-matrix.
Moreover, one has potentially large contributions also due to the exchange of the six sub-dominantly coupled heavy quasi-Dirac states, that can lie at the TeV scale, or even lower. 

In our dark linear seesaw, charged lepton flavor violation can also be mediated by the charged scalar $\eta ^{\pm }$ and the dark fermions $F_{i}$ through the couplings $Y^{(\eta)}$, as shown in Fig.~\ref{fig:figmuegFeyn}. 
This second contribution is especially interesting as the same dark sector Yukawa couplings $Y^{(\eta)}$ generating neutrino masses radiatively via the linear seesaw can
also give rise to charged lepton flavor violation. 
The Feynman diagrams for these two contributions are shown in Fig.~\ref{fig:figmuegFeyn}.

The total branching ratio for the process $\mu \rightarrow e \gamma $ thus takes the form \cite{Langacker:1988up,Lavoura:2003xp,Hue:2017lak},
\begin{eqnarray}
Br\left( \mu \to e \gamma \right) &=& \frac{3\left( 4\pi \right)\alpha _{em}}{4G_{F}^{2}} \left\vert \sqrt{\frac{%
\alpha _{W}^{2}s_{W} }{m_{W}^{4}}}%
\sum_{k=1}^{9}K_{2k}^{\ast }K_{1k}G_{F}\left( \frac{M_{k}^{2}}{m_{W}^{2}}%
\right) \right.  \notag \\
&&\left. + 
\sum_{k=1}^{3}\frac{Y_{2k}^{\left( \eta \right) }Y_{2k}^{\left( \eta \right)
}}{2 m_{\eta ^{\pm }}^{2}}G_{\eta ^{\pm }}\left( 
\frac{m_{F_{k}}^{2}}{m_{\eta ^{\pm }}^{2}}\right) \right\vert ^{2},\hspace{%
0.5cm}\hspace{0.5cm}\hspace{0.5cm}\hspace{0.5cm}  \label{Brmutoegamma} \\
&&\text{with,}  \notag \\
G_{F}\left( x\right) &=&\frac{10-43x+78x^{2}-49x^{3}+18x^{3}\ln x+4x^{4}}{%
12\left( 1-x\right) ^{4}}, \\
G_{\eta ^{\pm }}\left( x\right) &=&\frac{1-6x+3x^{2}+2x^{3}-6x^{2}\ln x}{%
6\left( 1-x\right) ^{4}}.\;
\end{eqnarray} 
In Eqn. (\ref{Brmutoegamma}), the matrix $K$ is the $3\times 9$ rectangular mixing matrix describing the CC weak interaction
and includes the exchange of the three light active neutrinos with $k=1,2,3$ as well as the six mediators, with $k=4,5,..9$.
As mentioned earlier these form three quasi-Dirac heavy-neutrino pairs.  

{The complete form of the lepton mixing matrix $K$ is given by:  
\begin{equation}
K=\left( K_{L},K_{H}\right) ,
\end{equation}
where $K_{L}$ and $K_{H}$ are $3\times 3$ and $3\times 6$ matrices,
respectively. These submatrices take the form:
\begin{eqnarray}
K_{L} &=&\left( 1_{3\times 3}-\frac{1}{2}m_{D}\left( M^{-1}\right)
^{T}M^{-1}m_{D}^{\dagger }\right) U_{lep}=\left( 1_{3\times 3}-\frac{1}{2}%
VV^{\dagger }\right) U_{lep}\,, \\
K_{H} &=&\left( -\frac{i}{\sqrt{2}}V,\frac{1}{\sqrt{2}}V\right) ,\hspace{1cm}%
V=m_{D}\left( M^{-1}\right) ^{T}.
\end{eqnarray}}
\begin{figure*}[tbh]
  \includegraphics[scale=0.78]{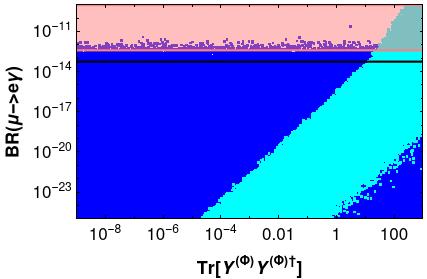}
  \includegraphics[scale=0.78]{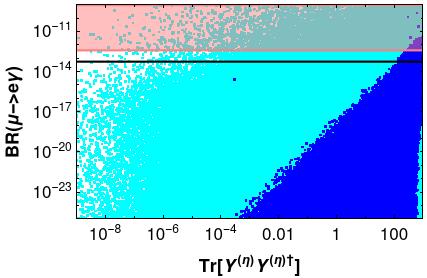}
\caption{
  Correlations of $Br(\protect\mu \to e \protect\gamma)$ against the Tr$[Y^{(\Phi)} {Y^{(\Phi)}}^\dag]$ and  Tr$[Y^{(\eta)} {Y^{(\eta)}}^\dag]$.
  The cyan points corrspond to the CC contribution, whereas the blue points show the dark-sector contribution.}
\label{fig:figmueg}
\end{figure*}

\begin{table}[h]
\setlength\tabcolsep{0.25cm} \centering%
\begin{tabular}{|c||c|c|c|c|c||c|c|c|}
\hline
Parameters & $Y_{ij}^\eta$ & $ Y_{ij}^\xi = y\delta_{ij}$ & $m_{F_i}$ & $M_{ij} =M_N \delta_{ij}$ & $m_{D_1,D_2,D_{A_1},D_{A_2},\eta^{\pm}} $  & $\theta_{D}$ & $a,b,c$ \\ \hline
Range  & $[10^{-10}, 4\pi]$ & $[10^{-16},4\pi]$ & 
$[200,5000]$ GeV & $[200,5000]$ GeV  & $[200,5000]$ GeV  &0.01 & [-20,20] \\ \hline
\end{tabular}%
\caption{The sampling region used in generating the plots of Fig.~\ref{fig:figmueg}.}
\label{tab:scan}
\end{table}

In Fig.~\ref{fig:figmueg}, we present the correlations of $Br(\protect\mu \to e \protect\gamma)$ against Tr$[Y^{(\Phi)} {Y^{(\Phi)}}^\dag]$ (left) and Tr$[Y^{(\eta)} {Y^{(\eta)}}^\dag]$ (right). 
In order to optimize our parameter scan to generate these figures, ensuring that only viable solutions consistent with neutrino oscillation data are included, it is useful to use the analytical approximation in Eq.~(\ref{mdeqn}).
Note however, that in presenting the numerical results we use the exact expressions for the diagonalization matrices.

In generating Fig.~\ref{fig:figmueg}, the neutrino oscillation parameters are varied in their $3\sigma $ ranges~\cite{deSalas:2020pgw},
the parameters $a$, $b$ and $c$ are varied in the range $ [-20,20]$ and the couplings $Y_{ij}^{(\eta)}$ are varied up to $4\pi$. 
For simplicity we took the heavy neutrinos as degenerate, varying their masses in the range $[200,5000]$ GeV.  
Concerning the dark sector parameters, $Y_{ij}^{(\xi)}$ is taken as $y \delta_{ij}$ with $y$ varied up to $4\pi$.
The masses of the dark fermions $F_i$ and the scalar masses are varied in the range $[200,5000]$ GeV, { while the
scalar mixing angle $\theta _{D}$ is fixed to be 0.01 implying $\theta _{D_A} = -0.01$.} The sampling region is summarized in table \ref{tab:scan}. 

In Fig.~\ref{fig:figmueg}, the cyan and the blue points show the CC and the dark-sector contributions to $Br(\protect\mu \to e \protect\gamma)$, respectively. 
The horizontal pink-shaded region corresponds to the current bound~\cite{MEG:2016leq}, $Br(\protect\mu \to e \protect\gamma) < 4.2 \times 10^{-13}$ as obtained from the MEG experiment,
whereas the black line corresponds to the projected future sensitivity of $6 \times 10^{-14}$ for MEG-II~\cite{Baldini:2013ke,Meucci:2022qbh}. 
From the expression for the branching ratio, one can see that the CC contribution depends on $Y^{(\Phi)}$, whereas the dark sector contribution depends only on $Y^{(\eta)}$.
This correlation can also be seen from Fig.~\ref{fig:figmueg}. 
We find that even if the CC contribution is low (in the regions of small $Y^{(\Phi)}$), for large values of $Y^{(\eta)}$ the dark sector contribution can take values as large as the existing limit.
Part of this parameter space will be probed by MEG-II.

\subsection{Dark matter phenomenology}
\label{sec:dark-matt-phen}

In this section we discuss the implications of our model for dark matter.
Due to the remnant $\mathcal{Z}_{2}$ symmetry arising from the spontaneous breaking of the global $U\left( 1\right) _{\cal L }$ lepton number
symmetry, our model will have a stable dark matter candidate, which we call the LDP.

\subsubsection{Fermionic dark matter}

\begin{figure*}[!h]
\begin{center}
  \includegraphics[scale=0.3,height=4cm]{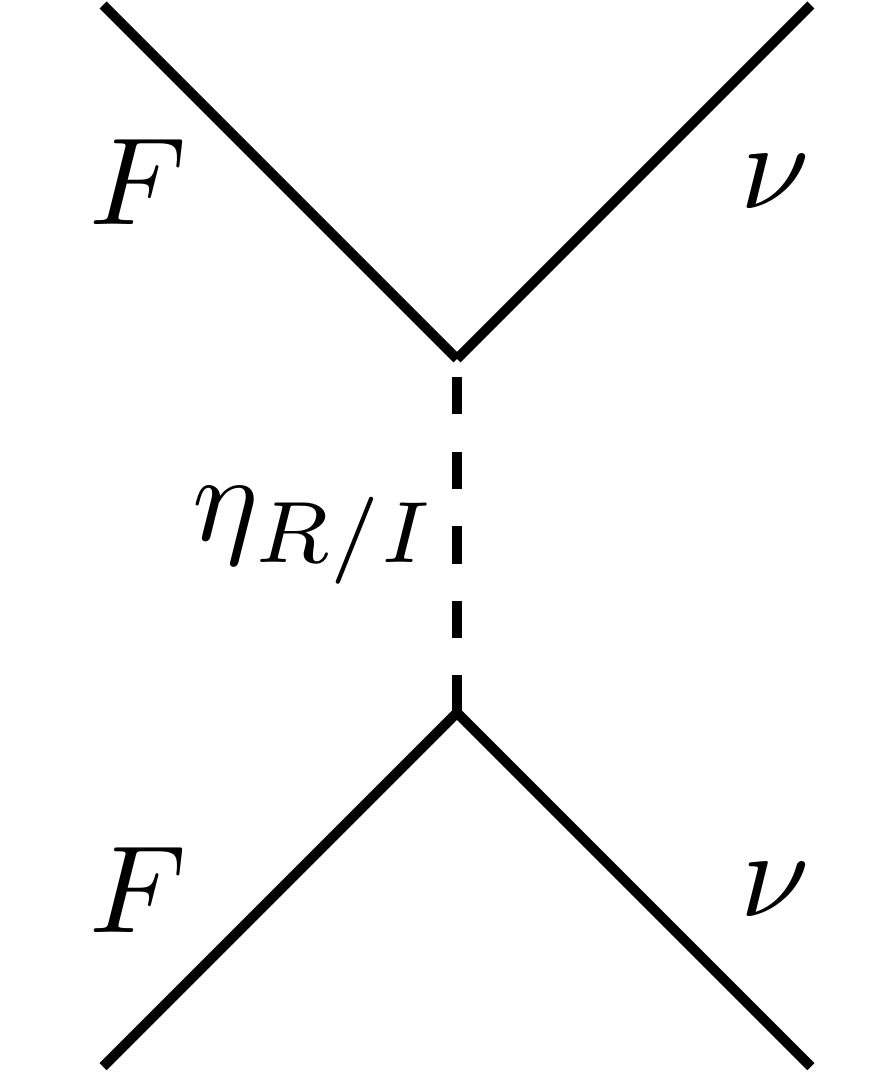}
\end{center}
\caption{
  Annihilation of a pair of fermionic dark matter candidates into a pair of active neutrinos.}
\label{figfermionDM}
\end{figure*}

 As a warm up, we start by considering a simple scenario in which the LDP is fermionic, i.e. the lightest of the heavy Majorana fermions $F_{i}$ ($i=1,2,3$).
It can annihilate into a pair of SM active neutrinos via the $t$-channel exchange of the
CP-even and CP-odd parts of the neutral component of the dark scalar doublet $\eta$, as shown in the Feynman diagram of Fig.~\ref{figfermionDM}. \black

In this case, the thermally-averaged annihilation cross section is given by~\cite{Bernal:2017xat}, 
\begin{equation}
  \label{eq:relic-fermion}
  \vev{ \sigma v} \simeq \frac{9\left(Y^{\left(\eta\right)}_{11}\right)^{4}}{32\pi }\frac{m_{F_1}^{2}\left( 2m_{F_1}^{2}+m_{D_1}^{2}+m_{D _{A_1}}^{2}\right)^2 }{\left( m_{F_1}^{2}+m_{D
_{1}}^{2}\right) ^{2}\left( m_{F_1}^{2}+m_{D_{A_1}}^{2}\right) ^{2}},
\end{equation}
where we have assumed $F_1$ to be the lightest among $F_i$.
Here $Y^{\left(\eta\right)}_{11}$ is the Yukawa coupling with the dark scalar doublet $\eta $.
From the previous relation, we find the following estimate for the DM relic abundance \cite{ParticleDataGroup:2022pth},  
\begin{equation} 
\frac{\Omega _{DM}h^{2}}{0.12}=\frac{0.1pb}{0.12\vev{\sigma v}}=\frac{0.1pb}{0.12%
}\left[ \frac{9\left(Y^{\left(\eta\right)}_{11}\right)^{4}}{32\pi }\frac{m_{F_1}^{2}\left( 2m_{F_1}^{2}+m_{D_1}^{2}+m_{D _{A_1}}^{2}\right)^2 }{\left( m_{F_1}^{2}+m_{D
_{1}}^{2}\right) ^{2}\left( m_{F_1}^{2}+m_{D_{A_1}}^{2}\right) ^{2}}\right]
^{-1},
\end{equation}
which in turn can reproduce the observed DM relic abundance of~\cite{Planck:2018vyg},
\begin{equation}
\Omega _{DM}h^{2}= 0.1200\pm 0.0012
\label{Omegavalue}
\end{equation}

\begin{figure}[tbh]
\centering
\includegraphics[width=0.5\textwidth,height=5cm]{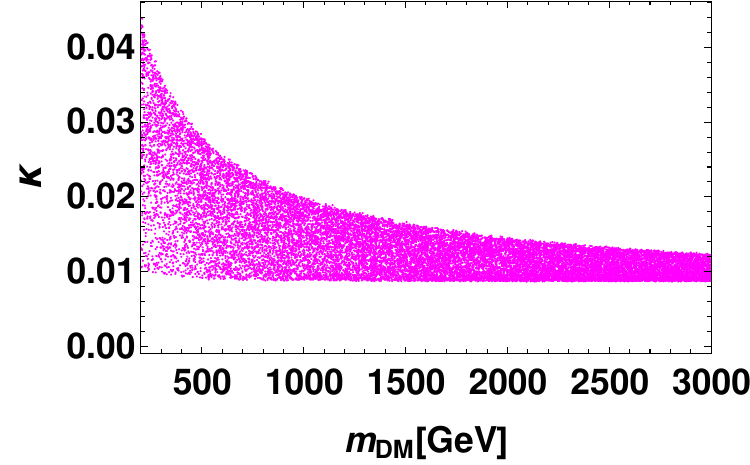}
\caption{ Allowed region in the $m_{DM} - \kappa $ plane that reproduces the correct fermionic dark matter relic density.
  Here $m_{DM}$ is the mass of the lightest dark fermion, i.e. $m_{F_1}$, and $\kappa$ is the corresponding Yukawa coupling, $Y^{\left(\eta\right)}_{11}$. } 
\label{DMplotfermion}
\end{figure}

In Fig.~\ref{DMplotfermion} we present the allowed parameter space in the $m_{F_1} - Y_{11}^{(\eta )} $ plane that reproduces the measured values of the dark matter relic abundance.  
In generating this figure, the masses of the dark scalars $m_{D_1}$ and $m_{D_{A_1}}$ are both varied in the range 0.2-5 TeV.
From the figure, one can see that the allowed range of the Yukawa couplings for lower values of the dark matter mass is broader.

However, since the fermionic dark matter $F_1$ couples only to the dark scalars, this scenario can not be probed by direct detection techniques based on nuclear recoil measurements.
For this reason we now move to an alternative interesting scenario where one of the neutral scalars of the dark sector is the LDP, and hence the DM candidate. 

\subsubsection{Scalar dark matter} \black

Scenarios with a scalar DM candidate have been well studied within the framework of generalized scotogenic or inert doublet
models~\cite{LopezHonorez:2006gr,Hirsch:2013ola,Restrepo:2019ilz,Avila:2019hhv,Kang:2019sab,CarcamoHernandez:2020ehn,Hernandez:2021zje,Hernandez:2021iss,CarcamoHernandez:2022vjk}. 
This DM possibility can arise in our present model by assuming that the LDP is the lightest among the neutral scalar particles $D_{1}$, $D_{2}$, $D_{A_1}$, $D_{A_2}$,
and lighter than the heavy neutral Majorana fermions $F_i$. 
In the following discussion, we take small doublet-singlet mixing angles so that $D_{A_1}$, taken as the DM candidate,
  is mainly the imaginary part of the neutral component of the dark doublet $\eta$.
Such a scalar DM candidate would scatter off a nuclear target through Higgs boson exchange in the $t$-channel.
This gives rise to a direct Higgs portal dark-matter detection mechanism that can be used to probe the coupling parameter characterizing the $H_1^{2}D_{A_1}^{2}$ interaction.

\begin{figure*}[tbh]
\begin{center}
  \includegraphics[scale=0.38]{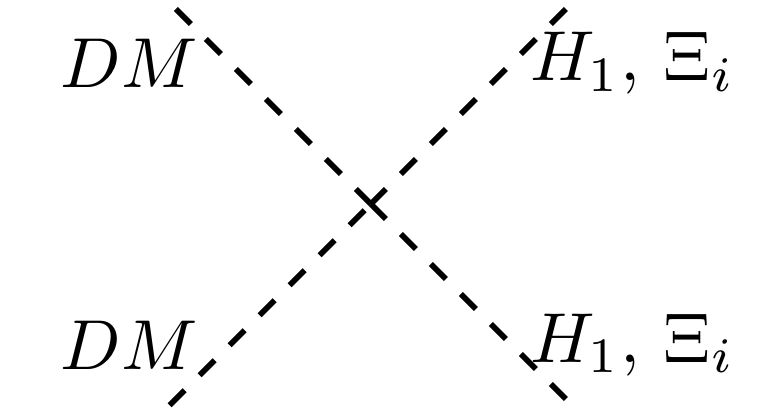}
  \includegraphics[scale=0.38]{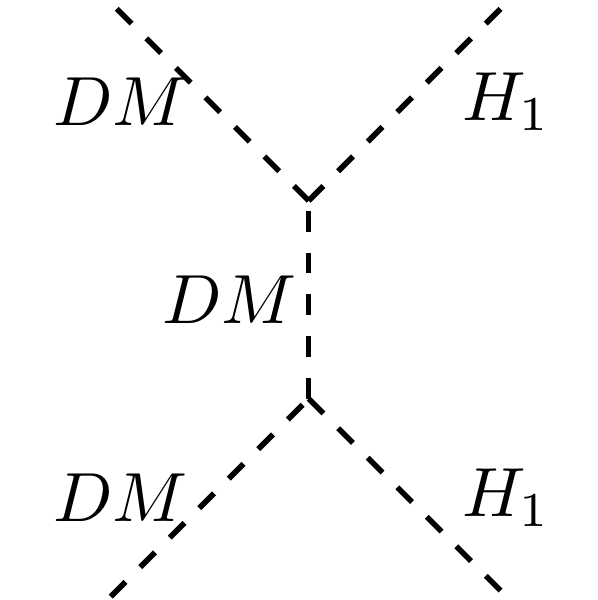}
  \includegraphics[scale=0.31]{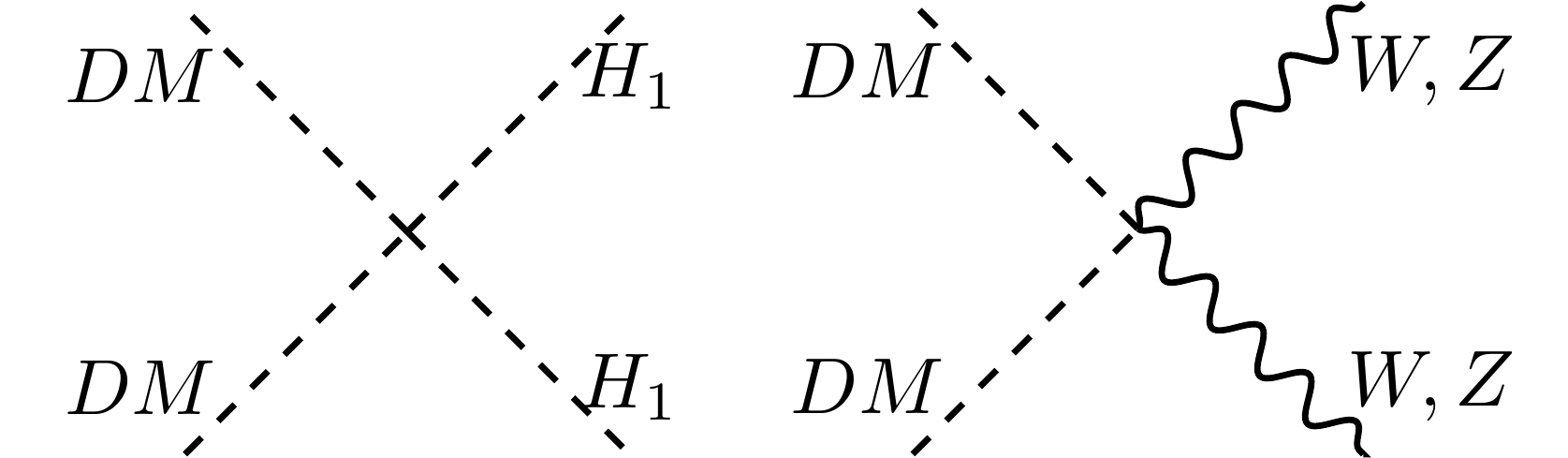}
\end{center}
\caption{
 Dominant Feynman diagrams contributing to dark-matter annihilation for the benchmark considered.}
\label{fig:figDMscalar}
\end{figure*}

\begin{figure}[tbh]
\centering
\includegraphics[width=0.5\textwidth,height=5cm]{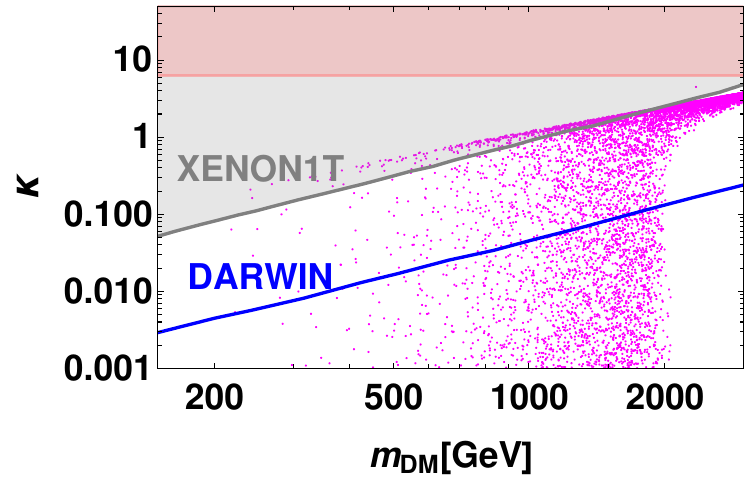}
\caption{ Allowed $m_{DM} - \kappa $ region that reproduces the correct DM relic density (magenta points) for the scalar dark matter.
  Here, $\kappa$ is the DM coupling to the SM Higgs boson. The pink band is disfavored by perturbativity.
  The gray region is excluded by XENON1T~\cite{XENON:2018voc} while the blue line corresponds to the sensitivity of DARWIN~\cite{DARWIN:2016hyl}.}
\label{DMplot}
\end{figure}

We assume that the main coannihilation channels for the DM candidate lead to a pair of SM particles as well as charged and the neutral components of the scalar triplet,
as shown in Fig.~\ref{fig:figDMscalar}. In this benchmark scenario, we take all dark sector quartic couplings except $\lambda_{6,7,11}$ to be small.
Thus, the dominant contributions to the DM coannihilation are shown in Fig.~\ref{fig:figDMscalar}. 
The corresponding annihilation cross sections are given as \cite{Hambye:2009pw,Bhattacharya:2016ysw},
\begin{eqnarray}
v_{rel}\sigma \left( D_{A_{1}}D_{A_{1}}\to WW\right)  &=&\frac{1}{8\pi m_{DM}\sqrt{s}}\frac{g^4}{4}\Big(1+ \frac{m_{DM^4}}{m_W^4} \Big( \frac{\Delta m^2 + \kappa v_\Phi^2}{m_{DM}^2} \Big)^2\Big)\sqrt{1-\frac{4 m_W^2}{s}}, \\
v_{rel}\sigma \left( D_{A_{1}}D_{A_{1}}\to ZZ\right)  &=&\frac{1}{16\pi m_{DM}\sqrt{s}}\frac{g^2}{4 c_w^4}\Big(1+ \frac{m_{DM^4}}{m_Z^4} \Big( \frac{\Delta m^2 + \kappa v_\Phi^2}{m_{DM}^2} \Big)^2\Big)\sqrt{1-\frac{4 m_Z^2}{s}}, \\
v_{rel}\sigma \left( D_{A_{1}}D_{A_{1}}\to q\overline{q}\right)  &=&
\frac{N_{c}\kappa^{2}m_{q}^{2}}{16\pi }\frac{\sqrt{\left( 1-\frac{4m_{f}^{2}}{s}\right) ^{3}}}{\left( s-m_{H_1}^{2}\right) ^{2}+m_{H_1}^{2}\Gamma_{H_1}^{2}}, \\
v_{rel}\sigma \left( D_{A_{1}}D_{A_{1}}\to H_1 H_1\right)  &=&\frac{\kappa^{2}}{64\pi s}\left( 1+\frac{3m_{H_1}^{2}}{s-m_{H_1}^{2}}-\frac{2\lambda
v^{2}}{s-2m_{H_1}^{2}}\right) ^{2}\sqrt{1-\frac{4m_{H_1}^{2}}{s}},\\
v_{rel}\sigma \left( D_{A_{1}}D_{A_{1}}\to \Xi_i \Xi_i\right)  &=&\frac{\lambda_{11} ^{2}}{64\pi s}\sqrt{1-\frac{4m_{\Xi_i}^{2}}{s}},
\end{eqnarray}%
where $\kappa = \frac{\lambda_6 + \lambda_7}{4}$, $m_{DM} = m_{D_1}$, $\Delta m^2 = m_{D_{A_1}}^2 - m_{D_1}^2$, the mass splitting between the CP even and the odd parts of the neutral component of $\eta$,
$\sqrt{s}$ is the centre-of-mass energy, $N_{c}=3$ stands for the color factor, $m_{H_1}=125.7$ GeV and $\Gamma _{H_1}$ is the total decay width of the SM Higgs boson which is equal to 4.1 MeV. 
 Here, the final state represented as $\Xi_i \Xi_i$ stands for $\Xi_1^+\Xi_1^-$, $\Xi_2^+\Xi_2^-$, $\Xi_I^0\Xi_I^0$, or $\Xi_R^0 \Xi_R^0$.
From this, the dark matter relic abundance in the present Universe is estimated as follows (c.f. Ref.~\cite{ParticleDataGroup:2022pth,Edsjo:1997bg}),
\begin{equation}
\Omega h^{2} = \frac{0.1~~\textrm{pb}}{\langle \sigma v \rangle },\,%
\hspace{1cm}\langle \sigma v \rangle =\frac{A}{n_{eq}^{2}},
\end{equation}%
where $\langle \sigma v \rangle $ is the thermally averaged
annihilation cross section, $A$ is the total annihilation rate per unit
volume at temperature $T$ and $n_{eq}$ is the equilibrium value of the
particle density, which are given as~\cite{Edsjo:1997bg},
\begin{eqnarray}
A &=&\frac{T}{32\pi ^{4}}\int_{4m_{D_{A_{1}}}^{2}}^{\infty}\sum_{p=W,Z,t,b,H_1,\Xi_i} g_{p}^{2}\frac{s\sqrt{s-4m_{D _{A_1}}^{2}}}{2}%
v_{rel}\sigma \left( D_{A_{1}}D_{A_{1}}\rightarrow SM SM\right)
K_{1}\left( \frac{\sqrt{s}}{T}\right) ds,  \notag \\
n_{eq} &=&\frac{T}{2\pi ^{2}}\sum_{p=W,Z,t,b,H_1,\Xi_i}g_{p}m_{D_{A_1}}^{2}K_{2}\left( \frac{m_{D_{A_1}}}{T}\right) 
\end{eqnarray}
with $K_{1}$ and $K_{2}$ being the modified Bessel functions of the second kind of order 1 and 2, respectively \cite{Edsjo:1997bg}.  
For the relic density calculation, we take $T=m_{m_{A_1}}/20$ as in Ref.~\cite{Edsjo:1997bg}, which corresponds to a typical freeze-out temperature.  
The DM relic density thus determined should match the required value as in Eqn.~\ref{Omegavalue}.
 In Fig.~\ref{DMplot}, we display the allowed parameter space (magenta points) in the $m_{DM}-\kappa$ plane that reproduces the correct dark matter relic abundance.
In calculating the dark matter relic density, we have considered the annihilation of the DM into $WW$, $ZZ$, $H_1 H_1$, $t\overline{t}$, $b\overline{b}$, and the components of the triplet scalar, which are the dominant channels.  The masses of the charged and neutral components of $\Xi$ are varied in the range 0.2-5 TeV and the quartic coupling $\lambda_{11}$ is varied in the range $0-4\pi$.
 The pink band is disfavored by perturbativity whereas the gray shaded region is disfavored by the bounds from XENON1T~\cite{XENON:2018voc}. The region above the blue line will be probed by the experiment DARWIN~\cite{DARWIN:2016hyl}.  

\subsection{Collider experiments}  
\label{sec:colliders}

In this subsection we briefly discuss the collider signatures associated to the dark matter candidates in our model.  
Due to the remnant $\mathcal{Z}_{2}$ symmetry, the scalar and fermionic dark matter candidates will be produced in pairs. 
Because of the small mixing between $\eta^0_{R}$ and $\xi^0_{R}$ as well as $\eta^0_{I}$ and $\xi^0_{I}$, we can take them aproximatelly as the mass eigenstates.
The neutral components of the dark doublet can be produced in pairs via the Drell-Yan mechanism mediated by the $Z$-boson or through vector boson fusion. This will correspond to a final state of 2 jets plus missing energy at the collider.
Detailed studies of collider signatures arising from pair production of neutral components of the dark doublet via vector boson fusion are provided in~\cite{Dutta:2017lny}.

\begin{figure}[tbh]
\centering
\includegraphics[width=0.4\textwidth,height=4.5cm]{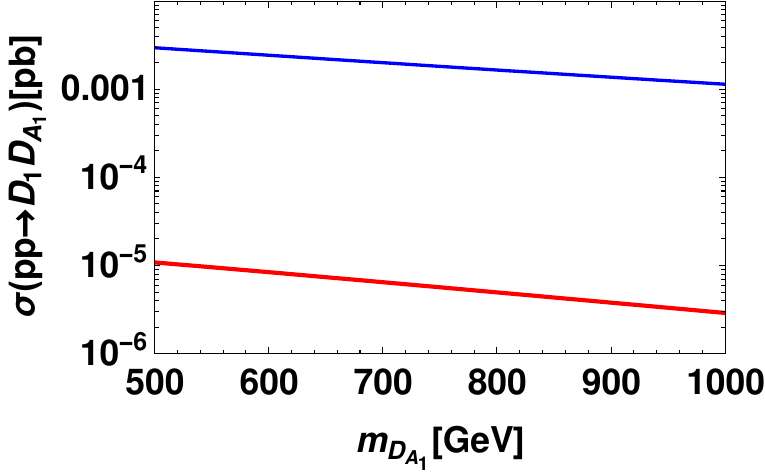}
\caption{
  Total cross section for the CP-even and CP-odd scalar dark-matter production via the Drell-Yan mechanism at a proton-proton collider for $\protect\sqrt{s}=14$
TeV (red line) and $\protect\sqrt{s}=100$ TeV (blue line) as a function of the CP-odd scalar dark matter mass $m_{D_{A_1}}$.}
\label{fig:pptoD1Da1}
\end{figure}

Fig.~\ref{fig:pptoD1Da1} displays the total cross section for the pair production of $\eta^0_R$ and $\eta_0^I$ via the Drell-Yan mechanism at a proton-proton collider for $\protect\sqrt{s}=14$ TeV (red line)
and $\protect\sqrt{s}=100$ TeV (blue line) as a function of the CP-odd dark scalar mass $m_{D_{A_1}}$, taken to vary in the range from $500$ GeV up to $1.0$ TeV.
Here the mass of the CP even dark scalar, $m_{D_{1}}$ has been set to be equal to $1$ TeV.
As shown in Figure \ref{fig:pptoD1Da1}, the total cross section for the CP-even and CP-odd dark scalar production at the LHC via the Drell-Yan mechanism reaches values of the order of $10^{-5}$ pb
for $m_{D_{A_1}}$ equals to $0.5$ TeV, and decreases as $m_{D_{A_1}}$ takes larger values.

The total Drell Yan production cross section increases by two orders of magnitude when one consider a $100$ TeV proton-proton collider.
In this case, the cross section reaches a value as high as $3\times 10^{-3}$ pb when the CP-odd dark scalar mass is set to $0.5$ TeV. 
For the case of the fermionic DM candidate, the pair production of the charged components of the dark doublet through the Drell-Yan mechanism
and their subsequent decays can give rise to a signature with opposite-sign dileptons plus missing energy in the final state.
The observation of an excess of events of this opposite-sign dileptons final state configuration with respect to the SM background could provide support of this model at the LHC.
A detailed study of collider signatures lies beyond the scope of the present work and will be taken up elsewhere. 

\section{Summary and conclusions}
\label{sec:Conclusions}

In this work we have proposed a minimal model where neutrino mass generation arises at the one-loop level within the linear seesaw mechanism.
Lepton number violation is seeded by a dark sector involving three Majorana fermions and two types of dark scalars, one isodoublet and the other isosinglet under the \SM symmetry.
{The small neutrino masses arise from the spontaneous lepton number violation by a small Higgs triplet vacuum expectation value and involve the interplay of the one-loop dark sector seed with the linear seesaw mechanism.} 
Our multiplet choice prevents the appearance of unwanted tree-level mass terms that could contribute to neutrino masses, making them genuinely calculable. 

We have studied the predicted rates for charged lepton flavour violation, Figs.~\ref{fig:figmuegFeyn} and \ref{fig:figmueg}.
We briefly discuss the prospects for testing our framework with the results of current and future lepton flavour violation searches.
 We have also commented on the WIMP dark-matter phenomenology of our
  model.
  For example, Figs.~\ref{figfermionDM} and \ref{DMplotfermion} correspond to the case where lightest dark fermion acts as the DM candidate.
  On the other hand, Figs. \ref{fig:figDMscalar} and \ref{DMplot} are for the case in which the DM candidate is the lightest neutral scalar arising from the dark sector.
Finally we make some comments on possible collider implications, Fig.~\ref{fig:pptoD1Da1}. However, these would require a dedicated scrutiny outside the scope of this paper.

\textbf{Note added}

As this work was being completed we came to know that A. Batra, H. Camara and F. R. Joaquim have come up with an alternative realization of the same
idea~\cite{Batra:2023bqj}.
{Prompted by a discussion with them, we noticed and corrected an inconsistency in the first version of our paper.}  
We stress that, thanks to their different multiplet structures, the phenomenology of the two proposals is quite different. 

\acknowledgements 
\noindent

AECH has received funding from Chilean grants ANID-Chile FONDECYT 1210378, ANID PIA/APOYO AFB220004, ANID Programa Milenio code ICN2019$\_$044.
  V.K.N. is supported by ANID-Chile Fondecyt Postdoctoral grant 3220005.
  The work of J.V. is supported by the Spanish grants PID2020-113775GB-I00~(AEI/10.13039/501100011033) and Prometeo CIPROM/2021/054 (Generalitat Valenciana).

\newpage
\bibliographystyle{utphys}
\bibliography{bibliography}
\end{document}